\documentclass[twoside,12pt]{article}
\usepackage{epsfig}

\newcommand{\be}{\begin{equation}}
\newcommand{\ee}{\end{equation}}
\newcommand{\bea}{\begin{eqnarray}}
\newcommand{\eea}{\end{eqnarray}}

\begin{document}

\title{ \vspace{1cm} Study of the $\Sigma$(1385) and $\Lambda$(1405) resonances in \\
K$^+$ photoproduction
processes}
\author{Madeleine Soyeur$^{1}$, Matthias F.M. Lutz$^{2,3}$\\
\\
$^1$D\'{e}partement d'Astrophysique, de Physique des Particules,\\
de Physique Nucl\'{e}aire et de l'Instrumentation Associ\'{e}e,\\
Service de Physique Nucl\'{e}aire,
CEA/Saclay,\\
F-91191 Gif-sur-Yvette Cedex, France\\
$^2$GSI, Planckstrasse 1,  D-64291 Darmstadt, Germany\\
$^3$Institut f\"ur Kernphysik, TU Darmstadt,
   D-64289 Darmstadt, Germany\\}

\maketitle
\begin{abstract} The $\gamma p \rightarrow K^+ \pi^0 \Lambda$ and $\gamma p \rightarrow K^+ \pi \Sigma $
reactions are studied in the kinematic region where the $\pi^0 \Lambda$(1116) and $\pi\Sigma$(1192) pairs
originate dominantly from the decay of the $\Sigma$(1385) and $\Lambda$(1405) resonances.
We consider laboratory photon energies around 2 GeV, i.e. total center of mass energies
above the known resonance region. We compute the
t-channel kaon-exchange contribution to these
reactions using
$K^- p \rightarrow \pi^0 \Lambda$ and $K^- p \rightarrow \pi \Sigma$
amplitudes calculated in the framework of a chiral coupled-channel
effective field theory of meson-baryon scattering. We extract from the calculated cross section
the gauge-invariant double kaon pole term. We find this term to be large and
likely to drive significantly the $\gamma p \rightarrow K^+ \pi^0 \Lambda$
and $\gamma p \rightarrow K^+ \pi \Sigma $
reactions in the kinematics under investigation. Accurate measurements of t-distributions
for these processes, in progress or planned at ELSA and at SPring-8, are needed to confirm this
expectation and assess the possibility of studying antikaon-nucleon dynamics
just below threshold through these reactions.

\end{abstract}
\section{Introduction}
Antikaon-nucleon dynamics close to threshold ($\sqrt {s^{thresh}}$=1.43 GeV) appears rather complex.
The $\bar KN$ channel in that regime couples both to inelastic channels
($\Lambda \pi$ and $\Sigma \pi$) and to baryon resonances located below
and above threshold.
The baryon resonances present below threshold [$\Sigma$(1385) and $\Lambda$(1405)]
are strongly coupled to the hyperon-pion channels.
The $\Sigma^0(1385)$ decays primarily into the
$\pi^0\, \Lambda$ channel [($88\pm2)\,\%$] and less significantly [($12\pm2)\,\%$]
into the $\pi\, \Sigma$ channel \cite{Hagiwara}. The $\Lambda(1405)$ decays
entirely into the $\pi\, \Sigma$ channel \cite{Hagiwara}. We propose
to study the $K^- p \rightarrow \pi^0 \Lambda$ and $K^- p \rightarrow \pi \Sigma$
amplitudes below threshold, in the region where they are dominated by the
$\Sigma$(1385) and $\Lambda$(1405) resonances, by isolating a specific term
arising from $K^-$ t-channel exchange in the $\gamma p \rightarrow K^+ \pi^0 \Lambda$
and $\gamma p \rightarrow K^+ \pi \Sigma $ reactions induced by 2 GeV photons.
This term is characterized by its analytic structure, a double pole
linked to the $K^-$ propagator [1/$(m_K^2-t)^2$], fulfills
the requirement of gauge-invariance and seems to be important.

This work is largely motivated by a new generation of experiments in which
multihadron final states can be measured exclusively.
 Section 2 is devoted to a brief review of
these experimental projects.
We describe in Section 3 the calculation of the t-channel
K$^-$-exchange contribution
to the $\gamma p \rightarrow K^+ \pi^0 \Lambda$ and
$\gamma p \rightarrow K^+ \pi \Sigma $
reactions using the $K^- p \rightarrow \pi^0 \Lambda$
and $K^- p \rightarrow \pi \Sigma$ amplitudes obtained
in Ref. \cite{Lutz1}.
We extract the gauge-invariant double $K^-$ pole contribution for both processes.
A few numerical results for the
$\gamma \, p \rightarrow K^+ \, \pi^0 \,\Lambda(1116)$ and
$\gamma\, p \rightarrow K^+\, \pi\, \Sigma(1192) $ reactions based on the double K$^-$ pole term
are presented in Section 4. We conclude briefly in Section 5.

This talk is based on a recent paper \cite{Lutz2} to which we refer
for a more extensive account of our calculation and a broader presentation of the results.

\section{Experimental studies of the $\gamma \, p \rightarrow K^+ \, \pi^0 \,\Lambda(1116)$ and
$\gamma\, p \rightarrow K^+\, \pi\, \Sigma(1192) $ reactions}

Very little is presently known on the $\gamma \, p \rightarrow K^+ \, \pi^0 \,\Lambda(1116)$ and
$\gamma\, p \rightarrow K^+\, \pi\, \Sigma(1192) $ reactions induced by real photons
in the region where the invariant mass of the $\pi \Lambda$ or $\pi \Sigma$ pairs is
close to the $\Sigma$(1385) or $\Lambda$(1405) mass. The only
published data of relevance for these processes were obtained at DESY thirty years ago with
space-like photons, in electroproduction
experiments where the scattered electron and the produced K$^+$ were detected
in coincidence \cite{Azemoon}. The differential cross sections for the $e\, p\rightarrow e\, K^+\,Y$
reaction in these measurements characterize globally strangeness production processes
for hyperon mis\-sing masses ranging from 1.35 until 1.45 GeV. The
$\Sigma^0(1385)$ and $\Lambda(1405)$ channel cannot
be separated, so that the published cross sections are associated with the production of both resonances.
An interesting
trend of these data is that the t-dependence of the cross section, for given
photon energy and virtuality, seems to show a sharp drop as would be expected
if the dynamics were dominated by t-channel exchanges. It is best fitted
precisely by the double kaon pole [1/$(m_K^2-t)^2$] \cite{Lutz2}. The Mandelstam variable t
is defined as the square of the
4-momentum transfer from the proton target to the $\Sigma^0(1385)$ or $\Lambda(1405)$.
We caution that these electroproduction data have very large error bars and should be viewed as
an mere indication that our suggestion is not in contradiction with existing experimental information.

There are on the other hand ongoing and future programs pertaining directly
to the study of the $\gamma \, p \rightarrow K^+ \, \pi^0 \,\Lambda(1116)$ and
$\gamma\, p \rightarrow K^+\, \pi\, \Sigma(1192) $ reactions, where the resonances will be separated. The
$\Sigma^0(1385)$ decays into neutral ($\pi^0 \Lambda$) and charged ($\pi^+ \Sigma^-$,
$\pi^- \Sigma^+$) channels; its decay into a $\pi^0 \Sigma^0$ pair is forbidden.
The $\Lambda(1405)$ decays into all $\pi \Sigma$ channels ($\pi^+ \Sigma^-$,
$\pi^- \Sigma^+$, $\pi^0 \Sigma^0$).
The $\pi^0\, \Sigma^0$ decay is
therefore a unique signature of the $\Lambda(1405)$. The
measurement of the $\gamma \, p \rightarrow K^+ \, \pi^0 \,\Sigma^0$ reaction is intended at
ELSA (Bonn) where the $\pi^0\, \Sigma^0$ pair could be detected through
a multi-photon final state ($\pi^0\, \Sigma^0 \rightarrow \pi^0\,\Lambda(1116)\,\gamma \rightarrow
 \pi^0\,n \,\pi^0\,\gamma \rightarrow n \, 5\gamma$) with the Crystal Barrel \cite{Schmieden}.
Similarly the $\Sigma^0(1385)$ could be studied by its $\pi^0\, \Lambda$ decay into the
$n \, 4\gamma$ channel. The charged channels were also studied at ELSA with the
SAPHIR detector. The $\gamma \, p \rightarrow K^+ \, \pi^+ \,\Sigma^- \rightarrow
K^+ \, \pi^+ \,\pi^-\, n$ and the $\gamma \, p \rightarrow K^+ \, \pi^- \,\Sigma^+ \rightarrow
K^+ \, \pi^- \,\pi^+\, n$ reactions (where all charged hadrons are detected)
have been investigated in the energy
range 1.3$\,<E_\gamma<\,$2.6 GeV \cite{Schulday}.  The charged channels are also
presently studied at SPring-8/LEPS with incident photon
energies in the range 1.5$\,<E_\gamma^{Lab}<\,2.4$ GeV \cite{Ahn}. The analysis
of both the SAPHIR and LEPS data is in progress. These data are dominated by
effects arising from the presence of the $\Lambda(1405)$ resonance.

We expect these data to be quite accurate and to unravel the dynamics
underlying the $\Sigma^0(1385)$ and $\Lambda(1405)$ production for
photon laboratory energies ranging from threshold kinematics until
the 2 GeV region addressed in this work. Angular or t-distributions
in successive energy bins should carry that information.

\newpage

\section{Dynamics of the $\gamma \, p \rightarrow K^+ \, \pi^0 \,\Lambda(1116)$ and
$\gamma\, p \rightarrow K^+\, \pi\, \Sigma(1192) $ reactions}

The dynamics of the $\gamma \, p \rightarrow K^+ \, \pi^0 \,\Lambda(1116)$ and
$\gamma\, p \rightarrow K^+\, \pi\, \Sigma(1192) $ reactions
in the region where the invariant mass of the $\pi \Lambda$ or $\pi \Sigma$ pairs is
close to the $\Sigma$(1385) and $\Lambda$(1405) masses reflects the
nature of these resonances.

The $\Sigma$(1385) arises as a member of the ground state decuplet of baryons in the large N$_c$ limit of QCD.
 It is well described by quark models \cite{Isgur,Glozman} and has a Breit-Wigner
shape \cite{Barreiro}.

The $\Lambda(1405)$
is a complex baryonic state. Its mass, in particular the large
splitting between the $\Lambda_{1/2^-}(1405)$ and the $\Lambda_{3/2^-}(1520)$,
cannot be understood in the constituent quark model with residual quark-quark interactions
fitting the other low-lying baryonic states \cite{Isgur,Glozman}.
Sizeable $q^4\bar q$
components seem required \cite{Jaffe}. The $\Lambda(1405)$ has been
described as a bound kaon-nucleon system
\cite{Dalitz1,Siegel}, in particular
as a kaon-soliton bound state \cite{Callan,Blom}.
The $\bar K N$ nature of the $\Lambda(1405)$ was also inferred from the
SU(3) cloudy bag model description \cite{Veit1,Veit2}. Extensive studies
of the $\Lambda(1405)$ based on chiral Lagrangians
\cite{Lutz1,Garcia-Recio}
suggest that this resonance
is generated by meson-baryon interactions. The spectral shape of the $\Lambda(1405)$
departs from a Breit-Wigner \cite{Hemingway}.
It depends strongly on the initial and final states through which it is measured,
emphasizing the need for a full understanding of the coupling of the $\Lambda(1405)$
to its different decay channels.

We study the $\gamma p \rightarrow K^+ \pi^0 \Lambda$ and $\gamma p \rightarrow K^+ \pi \Sigma $
reactions with the idea of using future accurate data on these processes (mainly
t-distributions) to gain understanding of the $K^- p \rightarrow \pi^0 \Lambda$
and of the  $K^- p \rightarrow \pi \Sigma$ amplitudes below the $\bar K N$ threshold,
where they are dominated by the $\Sigma(1385)$ and $\Lambda(1405)$ re\-sonances.

This procedure requires that these reactions be significantly driven by the process
in which the ingoing photon dissociates
into a real K$^+$ and a virtual K$^-$, the off-shell K$^-$ scattering subsequently
off the proton target to produce the $\pi^0\, \Lambda$ or $\pi\, \Sigma$ pair.
The corresponding diagrams are displayed in Fig.~1.
Such dynamics would show in a sharp drop of the differential cross sections $d\sigma/dt$
with increasing $|t|$ (as suggested by the scarce data available \cite{Azemoon}).
This drop can have both a double pole component behaving like 1/$(m_K^2-t)^2$
and a single pole dependence going like 1/$(m_K^2-t)$.
The new data expected in the near future
should make it possible to separate these terms by
expressing the differential cross sections $d\sigma/dt$ as a superposition of
double and single K$^-$ pole terms and less singular contributions.
To support further our t-channel approach, it should be noted that there are no
reasons to expect significant
s-channel contributions to the $\gamma p \rightarrow K^+ \pi^0 \Lambda$ and
$\gamma p \rightarrow K^+ \pi \Sigma $ reactions at E$_\gamma\simeq2$ GeV.
The corresponding total center of mass energy is
$\sqrt s= 2.15$ GeV. There are no baryon resonances in that mass range
known to decay into the $K^+ \pi^0 \Lambda$ or $K^+ \pi \Sigma $ channels.
Effects from u-channel contributions are not expected at low t.
\begin{figure}[h]
\vglue 0.4 true cm
\noindent
\begin{center}
\mbox{\epsfig{file=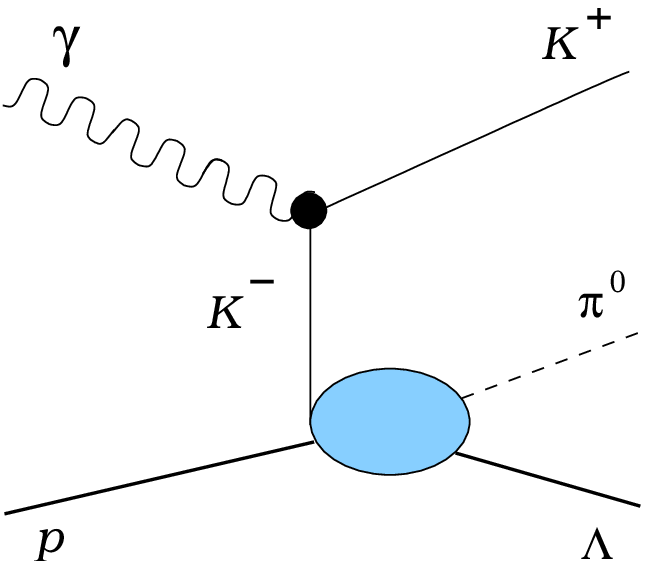, height=6 true cm}}
\end{center}
\begin{center}
\mbox{\epsfig{file=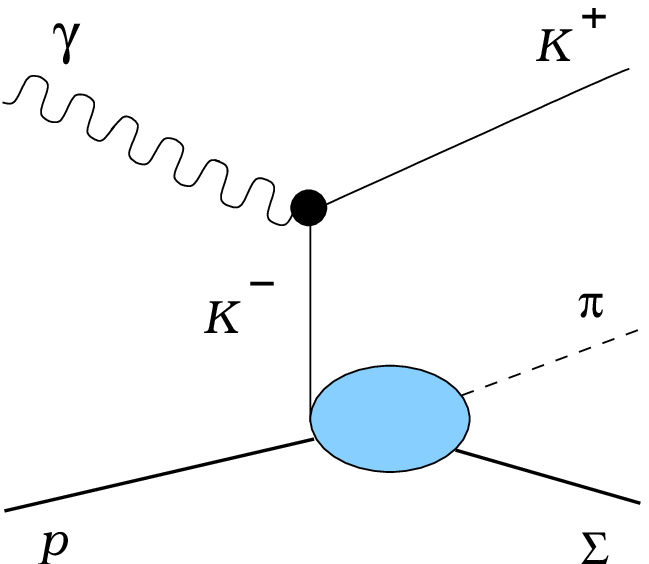, height=6 true cm}}
\end{center}
\vskip 0.4 true cm
\caption{K$^-$-exchange contribution to the
$\gamma \,p\rightarrow K^+ \pi^0 \Lambda$ amplitude (upper graph) and to the $\gamma \,p\rightarrow K^+ \pi \Sigma$ amplitude (lower graph). The $\pi \Sigma$
symbol stands for $\pi^- \Sigma^+$, $\pi^0 \Sigma^0 $ or $\pi^+ \Sigma^-$ in the latter case.}
\end{figure}

We have computed the K$^-$-exchange graphs of Fig. 1 using
the chiral coupled-channel approach of kaon-nucleon scattering
develo\-ped in Ref. \cite{Lutz1}. The different nature of the $\Sigma(1385)$
and of the $\Lambda(1405)$ resonances
is part of that picture. The baryon
resonances belonging to the large N$_c$ ground state baryon mutiplets
(hence the $\Sigma(1385)$) are introduced explicitly as fundamental fields
of the effective Lagrangian. The other baryon resonances (in
particular the $\Lambda(1405)$) are generated dynamically by meson-baryon
coupled-channel dynamics.
This effective field theory achieves an excellent description of the available data
on $K^-\,p$ elastic (direct and charge-exchange) and inelastic ($\pi^0\,\Lambda$,
$\pi^+\,\Sigma^-$, $\pi^0\,\Sigma^0$, $\pi^-\,\Sigma^+$) processes up to
laboratory $K^-$ momenta of the order of 500 MeV. The interest of the
present work is to offer the possibility of testing the amplitudes
below the $\bar K N$ threshold, in the region
where they are dominated by the $\Lambda$(1405) and the $\Sigma$(1385).
The specific spectral shape of these resonances is a particularly
meaningful prediction
of the description of Ref. \cite{Lutz1}
\newpage
We calculate the cross section for the $\gamma \, p \rightarrow K^+ \, \pi \,Y$,
where Y represents either the $\Lambda(1116)$ or the $\Sigma(1192)$. The 4-momenta
of the photon, the proton, the K$^+$, the pion and the hyperon are denoted
by $q$, $p$, $\bar q_K$, $\bar q_\pi$ and $\bar p_Y$ respectively.
The photon, proton and hyperon polarizations are indicated by the
symbols $\lambda_\gamma$, $\lambda$ and $\bar
\lambda_Y$. The total
cross section reads
\begin{eqnarray}
\sigma_{\gamma \,p\rightarrow K^+ \, \pi \,Y}&=&
\frac{1}{|{\vec v}_\gamma-{\vec v}_p|}\,\frac{1}{2\,q^0}\,\frac{m_p}{p^0}
\int \frac{d^3 \vec{\bar q}_K}{(2\pi)^3}\,\frac{1}{2\,{\bar q}\,^0_K}\,
\int \frac{d^3 \vec{\bar q}_\pi}{(2\pi)^3}\,\frac{1}{2\,{\bar q}\,^0_\pi}\,
\int \frac{d^3 \vec{\bar p}_Y}{(2\pi)^3}\,\frac{m_Y}{{\bar p}\,^0_Y}\,
\nonumber\\
&&(2\,\pi)^4\,\delta^4(q+p-{\bar q}_K-\bar q_\pi-\bar p_Y)\sum_{\lambda_\gamma,
\lambda ,\bar
\lambda_Y }\frac{1}{4}\,|M_{\gamma \,p\rightarrow K^+ \, \pi \,Y}|^2.
\label{eq1}
\end{eqnarray}
Factorizing  the full amplitude
$M_{\gamma \,p\rightarrow K^+ \, \pi \,Y}$ into the
photon-kaon vertex and the $K^-\,p\rightarrow \pi \,Y$ amplitude,
we can express the ${\gamma \,p\rightarrow K^+ \, \pi \,Y}$
cross section in terms of the $K^-\,p\rightarrow \pi \,Y$ cross section \cite{Lutz2}.
The latter is frame-independent and calculated for simplicity in the $K^-\,p$ center of mass.
It is useful to define the invariant mass $\sqrt{{\bar w}^2}$ of the final $\pi\,Y$ pair by
\begin{eqnarray}
{\bar w}^2 = (p+q-{\bar q}_K)^2
 = s+ m_K^2- 2\,\sqrt{s}\,\sqrt{m_K^2+{\vec{\bar q}_K}^{\,2}}.
\label{eq2}
\end{eqnarray}
\newpage
The expression for the total $K^-p \to \pi\,Y$
cross section is given by
\begin{eqnarray}
\sigma_{K^-\,p\rightarrow \pi \,Y} &=&
\frac {1} {\sqrt{{\bar w}^2} \, |\vec q_{K^-p}|} \, \frac{m_p}{2}\,
\int \frac{d^3 \vec{\bar q}_\pi}{(2\pi)^3}\,\frac{1}{2\,{\bar q}\,^0_\pi}\,
\int \frac{d^3 \vec{\bar p}_Y}{(2\pi)^3}\,\frac{m_Y}{{\bar p}\,^0_Y}\,
 \nonumber\\
&&\mkern 90 mu(2\,\pi)^4\,\delta^4({\bar w}-\bar q_\pi-\bar p_Y)\,
\frac{1}{2}\,
\sum_{\lambda ,\bar
\lambda_Y }\,|M_{K^-p \to \pi\,Y}|^2,
\label{eq3}
\end{eqnarray}
in which $q_{K^-p}$ is the $K^-$ momentum in the $K^-p$ center of mass,
\begin{eqnarray}
|\vec q_{K^-p}|^2=
\frac{1}{4\,{\bar w}^2}\, {\{{\bar w}^4-2\,{\bar w}^2\,(m_p^2+m_K^2)+(m_p^2-m_K^2)^2\}},
\label{eq4}
\end{eqnarray}
and the amplitudes $M_{K^-p \to \pi\,Y}$ are taken from Ref. \cite{Lutz1}.

In the energy range under consideration
(E$_\gamma\simeq2$ GeV), there are many possible diagrams contributing to the
$\gamma \,p\rightarrow K^+ \, \pi \,Y$ amplitudes and involving poorly known
couplings and hence large uncertainties.
The amplitudes $M_{\gamma \,p\rightarrow K^+ \, \pi \,Y}$
obtained by calculating the graphs of Fig. 1 are not gauge-invariant.
To obtain the full gauge-invariant amplitudes, a large class of diagrams of order $\alpha$ leading
to the same final state should be added.
We do not attempt to calculate these graphs and resort instead to the pole scheme
method \cite{Veltman}.
The idea of the method
is to decompose the amplitude according to its pole structure and to expand
it around the pole. To any order in perturbation theory, the residues
of the poles are gauge-invariant.
We apply this method to derive the gauge-invariant cross section
correspon\-ding to the double K$^-$-pole term.
The key point is that the graphs of Fig. 1 are the only process
which can contribute to the
double K$^-$-pole term. We will therefore decompose the corresponding cross section according to
its pole structure, keep only the double K$^-$-pole term and extract the
gauge-invariant cross section associated with that pole structure by
calculating the residue at the pole.

According to this procedure, the gauge-invariant cross section
corresponding to the double K$^-$-pole term reads
\begin{eqnarray}
\frac{d\sigma_{\gamma\,p\rightarrow K^+\,\pi \,Y}} {dt\, d{\bar w}^2} &=&
\frac{\alpha}{2\, \pi}\, \frac{({\bar w}^4-2\,{\bar w}^2\,(m_p^2+m_K^2)+(m_p^2-m_K^2)^2)^{1/2}}{(s-m_p^2)^2}\,
\nonumber\\
&&\mkern 150 mu\frac{m_K^2\,}{(t-m_K^2)^2}\,\sigma_{K^-\,p\rightarrow \pi \,Y}({\bar w}^2).
\label{eq5}
\end{eqnarray}
\par
We stress that the double pole term is the only one which can be determined
this way, because it does not get contributions from any other graph but the
t-channel kaon-exchange diagram.

In order to be able to extract the double pole term from accurate
t-distributions, it has to be reasonably large. We speculate so in
view of the numerical results displayed in the next section. As mentioned earlier
and discussed more thoroughly in Ref. \cite{Lutz2}, the double pole
behaviour is also compatible with the few data points available.

\section{Numerical results}

We show first the quantity $4\,|\vec q_{K^-p}|\sqrt{{\bar w}^2}$
$\sigma_{K^-\,p\rightarrow \pi \,Y}({\bar w}^2)$
as function of the total center of mass energy in the $K^- p$ system, renamed for clarity
$\sqrt{s_{K^-p}}$ ($\equiv{\sqrt{{\bar w}^2}}$). The interest of displaying our results
this way is to exhibit the behaviour of the ${K^-\,p\rightarrow \pi \,Y}$ cross section
across threshold. We recall that the $\bar K N$ threshold is at $\sqrt{s_{K^-p}}\,\approx\,$1.435 GeV.
We present for example in Fig. ~2 our predicted cross sections for the
${K^-\,p\rightarrow \pi^0 \,\Sigma^0}$ and
${K^-\,p\rightarrow \pi^0 \,\Lambda}$ reactions. They are compared to the data
available on these processes above threshold \cite{Mast1,Ciborowski,Armenteros,Humphrey}.
\par
\begin{figure}[h]
\noindent
\begin{center}
\mbox{\epsfig{file=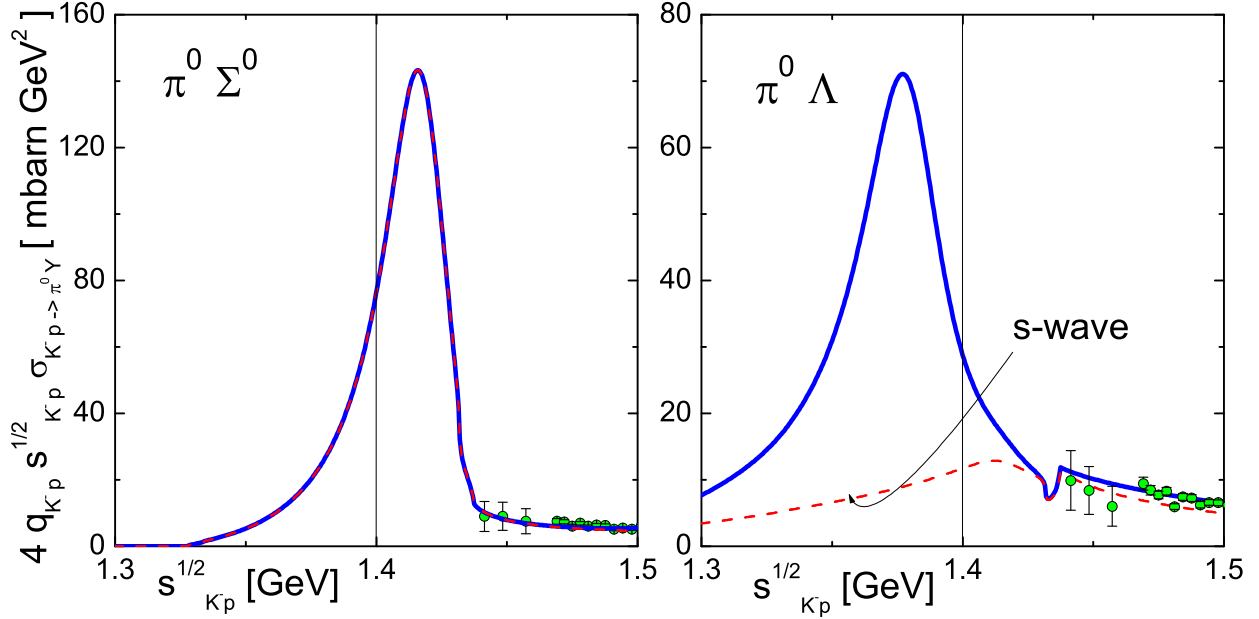, height=10 true cm}}
\end{center}
\caption{$K^- \, p \rightarrow \pi^0 \,\Sigma^0$ and $K^- \, p \rightarrow \pi^0 \,\Lambda$
cross sections below and above threshold. The length of the $K^-$ 3-momentum is defined
by Eq. (4) and $\sqrt{s_{K^-p}}$ ($\equiv{\sqrt{{\bar w}^2}}$) is the total center of mass energy
of the  $K^- \, p$ system. The dashed line represents the contribution from $K^- \, p$
relative s-wave only.
The data above threshold are from Refs. \cite{Mast1,Ciborowski,Armenteros,Humphrey}.}
\label{f2}
\end{figure}
\par
We recall that the $\pi^0 \,\Sigma^0$ channel reflects the $\Lambda$(1405)
and the $\pi^0 \,\Lambda$ channel the $\Sigma$(1385).
The properties of the spectral functions of the $\Sigma$(1385) and $\Lambda$(1405)
resonances are very apparent in Fig.~2. The shape of the resonant behaviour
of the ${K^-\,p\rightarrow \pi^0 \,\Lambda}$ cross section below threshold is quite symmetric and close
to a Breit-Wigner form. The s-wave contribution is small as expected for a
process dominated by a p-wave
resonance. In contrast, the spectral form of the ${K^-\,p\rightarrow \pi^0 \,\Sigma^0}$
cross section is asymmetric and largely given by
s-wave dynamics, reflecting the $\Lambda$(1405) dominance.

We display in Fig. 3 the
double kaon pole term contributions to the
differential cross sections for the $\gamma p \rightarrow K^+ \pi^0 \Lambda$ and
$\gamma p \rightarrow K^+ \pi \Sigma $ reactions
as functions of the $\pi\, Y$ total center of mass energy $\sqrt{s_{K^-p}}$ at
$E_\gamma=2.1$ GeV.
\par
\begin{figure}[h]
\noindent
\begin{center}
\mbox{\epsfig{file=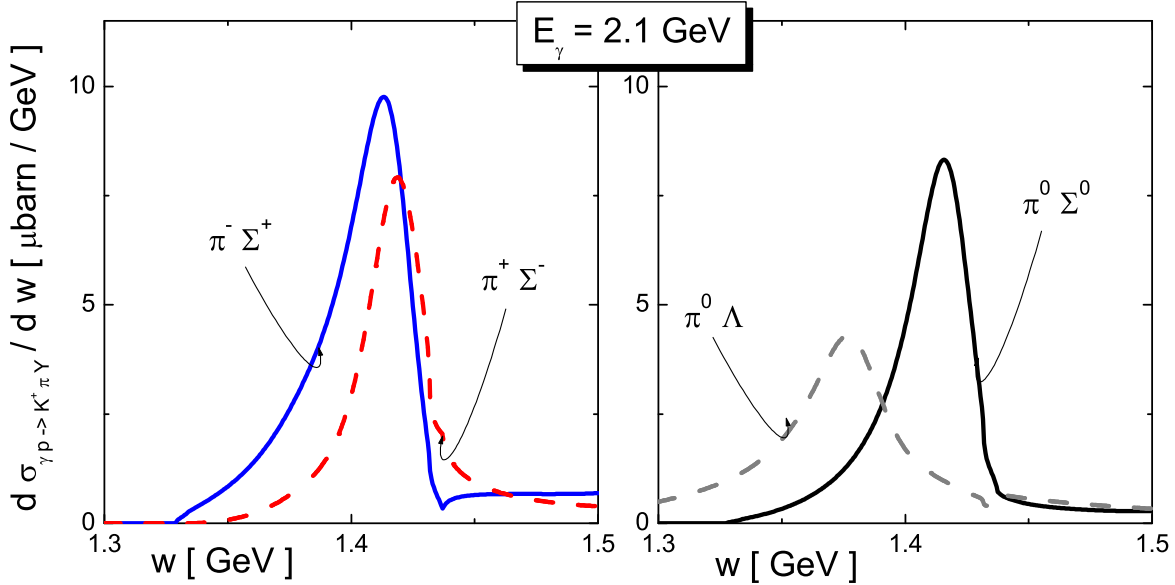, height=9.5 true cm}}
\end{center}
\caption{Double kaon pole term contribution to the
differential cross sections for the $\gamma p \rightarrow K^+ \pi^0 \Lambda$ and
$\gamma p \rightarrow K^+ \pi \Sigma $ reactions
as function of the $\pi\, Y$ total center of mass energy at $E_\gamma=2.1$ GeV
}
\label{f3}
\end{figure}
We see clearly the dynamical features discussed in commenting on
Fig. 3. It is also interesting to note the absolute values of the
double kaon pole cross sections.
They are large on the scale of what is expected from other theoretical approaches.
If we compare our results to the predictions of the model
of Ref. \cite{Nacher} at $E_\gamma=1.7$ GeV, we notice that our calculated
cross sections at that energy are roughly twice larger for the $\pi\,\Sigma$
channels \cite{Lutz2}.
It is not easy to trace the origin of this effect. Our gauge-invariant
double kaon pole term contains contributions which cannot be mapped easily
onto the Feynman diagrams computed in Ref. \cite{Nacher}.
The cross section we obtain for the $\pi^0\,\Lambda$ channel is about
an order of magnitude larger than the result displayed
in Ref. \cite{Nacher}. A substantial part of this effect should be
ascribed to the neglect of the $\Sigma$(1385) resonance in that work.
\newpage
\section{Conclusion}
We have studied the $\gamma p \rightarrow K^+ \pi^0 \Lambda$ and $\gamma p \rightarrow K^+ \pi \Sigma $
reactions in the kinematic region where the $\pi^0 \Lambda$(1116) and $\pi\Sigma$(1192) pairs
originate dominantly from the decay of the $\Sigma$(1385) and $\Lambda$(1405) resonances.
We focus on laboratory photon energies around 2 GeV, significantly above the threshold
for producing the $K^+\,\Sigma$(1385)
(E$^{thresh}_\gamma$=1.41 GeV)
and the $K^+\,\Lambda$(1405)
(E$^{thresh}_\gamma$=1.45 GeV)
final states. We have calculated the t-channel $K^-$-exchange contribution to these reactions
using the ${K^-\,p\rightarrow \pi \,Y}$ amplitudes of Ref. \cite{Lutz1}, which
have been shown to describe the data available at low kaon momentum. Based on the pole
structure of this contribution, we determined the gauge-invariant double kaon pole contribution to the
$\gamma p \rightarrow K^+ \pi Y$ cross sections by calculating the residue at the pole.
The relevance of our work stems from the advent of detector systems able
to measure exclusively multiparticle final states with great accuracy.
Three complementary experiments in the photon energy range considered in this paper
are planned with LEPS at SPring-8 \cite{Ahn}, SAPHIR at ELSA \cite{Schulday}
and  the Crystal Barrel at ELSA \cite{Schmieden},
dealing for the first two with the charged [$\pi^- \Sigma^+$ and $\pi^+ \Sigma^-$] channels
and for the latter with the neutral [$\pi^0 \Sigma^0$ and $\pi^0 \Lambda$] final states.
These accurate measurements
should make it possible to extract the contribution of the double kaon pole
and hence to study kaon-nucleon dynamics
below threshold.

\section{Acknowledgement}
We acknowledge very stimulating discussions with Takashi Nakano and Hartmut Schmieden.
One of us (M.S.) is grateful to the Erice School for supporting her participation
to a most fruitful course.

\end{document}